\documentclass[jcp, aip, floatfix, showpacs, reprint, citeautoscript]{revtex4-1}
\pdfoutput=1

\usepackage{amsmath}
\usepackage{graphicx}
\usepackage{dcolumn}
\usepackage[table]{xcolor}
\usepackage[version=3]{mhchem}
\usepackage{transparent}
\usepackage{gensymb}
\usepackage{etoolbox}

\def\maxfloatwidth{%
  \ifdim\columnwidth>246.0pt
  300.0pt  \else
  \columnwidth
  \fi
}

\newcommand{\trm}[1]{\textrm{#1}}

\newcommand{\etal}{\emph{et al.}}

\newcommand{\Eads}[2]{$E_{\trm{ads}}/\Delta H_{\trm{vap}}\ {#1}\ {#2}$}

\begin{document}

\date{\today}

\title{Molecular simulations of heterogeneous ice nucleation I:
  Controlling ice nucleation through surface hydrophilicity}

\author{Stephen J. Cox} 
\affiliation{Thomas Young Centre and Department of Chemistry,
  University College London, 20 Gordon Street, London, WC1H 0AJ, U.K.}
\affiliation{London Centre for Nanotechnology, 17--19 Gordon Street,
  London WC1H 0AH, U.K.}

\author{Shawn M. Kathmann} 
\affiliation{Physical Sciences Division, Pacific Northwest National
  Laboratory, Richland, Washington 99352, United States}

\author{Ben Slater} 
\affiliation{Thomas Young Centre and Department of Chemistry,
  University College London, 20 Gordon Street, London, WC1H 0AJ, U.K.}

\author{Angelos Michaelides} 
\email{angelos.michaelides@ucl.ac.uk}
\affiliation{Thomas Young Centre and Department of Chemistry,
  University College London, 20 Gordon Street, London, WC1H 0AJ, U.K.}
\affiliation{London Centre for Nanotechnology, 17--19 Gordon Street,
  London WC1H 0AH, U.K.}

\begin{abstract}

Ice formation is one of the most common and important processes on
Earth and almost always occurs at the surface of a material. A basic
understanding of how the physicochemical properties of a material's
surface affects its ability to form ice has remained elusive. Here we
use molecular dynamics simulations to directly probe heterogeneous ice
nucleation at an hexagonal surface of a nanoparticle of varying
hydrophilicity. Surprisingly, we find that structurally identical
surfaces can both inhibit and promote ice formation and analogous to a
chemical catalyst, it is found that an optimal interaction between the
surface and the water exists for promoting ice nucleation. We use our
microscopic understanding of the mechanism to design a modified
surface \emph{in silico} with enhanced ice nucleating ability.

\end{abstract}


\maketitle

Upon cooling, liquid water crystallizes into solid ice. Due to the
presence of a free energy barrier separating the liquid and
crystalline states, however, it is possible for liquid water to remain
in a metastable `supercooled' state to temperatures far below the
equilibrium melting temperature. Heterogeneous ice nucleation, that
is, ice nucleation in the presence of impurity particles such as
mineral dust, soot or certain types of bacteria, generally increases
the rate of ice nucleation and is the dominant process by which ice
forms in nature \cite{murray-review}. Recent work has argued that any
ice formation at temperatures above $-20$\celsius{} must necessarily
occur heterogeneously. \cite{sanz-homog:jacs} Empirically, a large
variance in the propensity of different materials to nucleate ice is
observed, and due to the importance of ice formation in e.g. the
climate sciences, much effort has been expended in identifying and
cataloging the effectiveness of different materials to nucleate ice
\cite{murray-review}. This has motivated many simulation studies of
heterogeneous ice nucleation in the presence of different surfaces,
including graphite \cite{molinero:het-jacs, molinero:het-jpca},
kaolinite \cite{FD:kaolinite, patey2008} and silver iodide
\cite{Patey:AgI, doye:AgI}. Despite the vast amount of research into
heterogeneous ice nucleation, major gaps in our knowledge still exist,
especially with regard to our understanding of the underlying chemical
physics; this is reflected in our inability to accurately predict a
material's ice nucleating efficiency and to answer seemingly simple
questions such as \textit{how does hydrophilicity affect the ice
  nucleation rate?} Not only is an understanding of the chemical
physics of heterogeneous ice nucleation needed to predict the ice
nucleating efficiency of existing materials \cite{bartels2013:nature,
  ipcc2013}, but it is also paramount for the rational design of new
materials to either promote or inhibit ice nucleation. Controlling ice
formation is desirable in a variety of fields, for example, in the
cryopreservation of cells, tissues and organs
\cite{gibson:cryopreservation}, the food and transport industries and
even as a potential means for climate control \cite{climate-eng1,
  climate-eng2}.

In contrast to fields such as chemical catalysis
\cite{norskov:NatChem-review} and materials design
\cite{curtarolo2013high}, there is currently no comprehensive set of
design principles in terms of molecular `descriptors' for making new
substances to control ice formation. Put more simply, we do not know
which are the relevant microscopic properties of a material that
determine its macroscopic ice nucleating efficiency. Often, the
so-called `requirements' for a good ice nucleating agent (INA) have
been discussed, such as the requirement for a good crystallographic
match to ice and the ability of water to chemically bond to the
surface of the particle (i.e. hydrophilicity) \cite{PK97}. Although
properties such as a good crystallographic match are important in
heterogeneous nucleation of some systems
\cite{frenkel:PRE:template,PhysRevLett:lattice-match:dynamicDFT}, such
criteria have neither served as a full set of guidelines to identify
good INAs \cite{murray-review}, nor have they aided the systematic
improvement of ice nucleation inhibitors or promoters. Experimentally
there is disagreement regarding the role of hydrophilicity. For
example, Alizadeh \etal{} \cite{alizadeh:hydrophilic} have found ice
nucleation to be slower on superhydrophobic surfaces, which they
attribute not only to a lower contact area between the water and the
surface, but also to a larger free energy barrier to nucleation. In
contrast, Li \etal{} \cite{li:hydrophobic, li:hydrophobic2} found ice
nucleation to be enhanced at hydrophobic modified silicon wafers
relative to their unmodified hydrophilic counterparts, which was
attributed to a faster dynamics of water at the hydrophobic
interface. Recently, Lupi \etal{} investigated the role of
hydrophiclity of graphitic surfaces using molecular dynamics
simulation \cite{molinero:het-jacs, molinero:het-jpca}: by varying the
hydrophilicity in different ways (by uniformly changing the
interaction of water with the surface or by introducing hydrophilic
species at the surface), they found that the ice nucleating efficacy
of the surface could either increase of decrease. Also, it is found on
kaolinite (a known hydrophilic INA) and platinum
\cite{kimmel:Pt-hydrophobic} that the most stable water overlayer can
inhibit the growth of subsequent water layers
\cite{xiaoliang2007,xiaoliang2008}. Furthermore, in the case of
requiring a good crystallographic match, evidence for ice-like
structures at surfaces is in general lacking \cite{NM:review} and
there are also instances where materials with a good crystallographic
match to ice are poor INAs \cite{sadtchenko2002,ewing:BaF2} (we note
that recent simulation studies \cite{Patey:AgI,doye:AgI} have found
the unreconstructed basal face of silver iodide to act as a template
for ice). We are therefore either faced with the prospect of relying
on experiments to determine the efficacy of INAs on a case-by-case
basis, or we can try and rationalize their behavior by elucidating the
underlying molecular processes that control heterogeneous ice
nucleation.

\begin{figure}[t]
  \centering
  \includegraphics[width=0.9\linewidth]{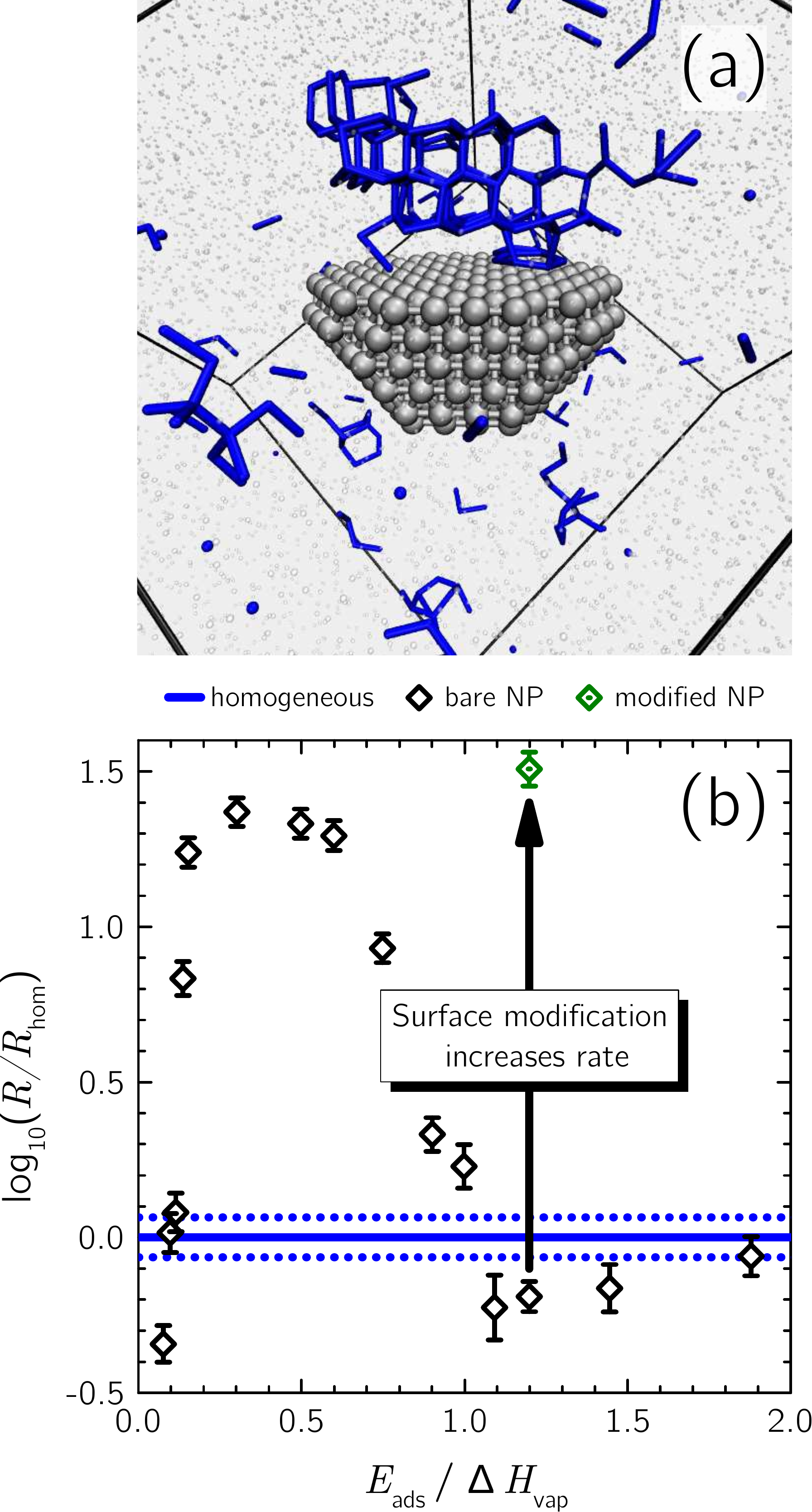}
  \caption{(color online) (a) Snapshot of a typical ice nucleation
    event on the NP. Ice-like molecules are colored blue and the NP is
    colored silver. The NP is totally immersed in water (liquid-like
    molecules are shown by gray dots). (b) Variation of the nucleation
    rate with the strength of the water-NP interaction
    $E_{\trm{ads}}$. As $E_{\trm{ads}}$ increases, so too does the
    hydrophilicity. The solid blue line indicates homogeneous
    nucleation (the dotted lines are an error estimate): data above
    and below this line indicates promotion and inhibition of ice
    nucleation by the NP, respectively. It can be seen that at weak
    and strong water-NP interaction strengths the NP inhibits
    nucleation, while at intermediate interaction strengths
    (\Eads{\approx}{0.15\trm{--}0.6}), the NP strongly promotes
    nucleation.}
  \label{fig:plots}
\end{figure}

\begin{figure*}[t]
  \centering
  \includegraphics[width=0.9\linewidth]{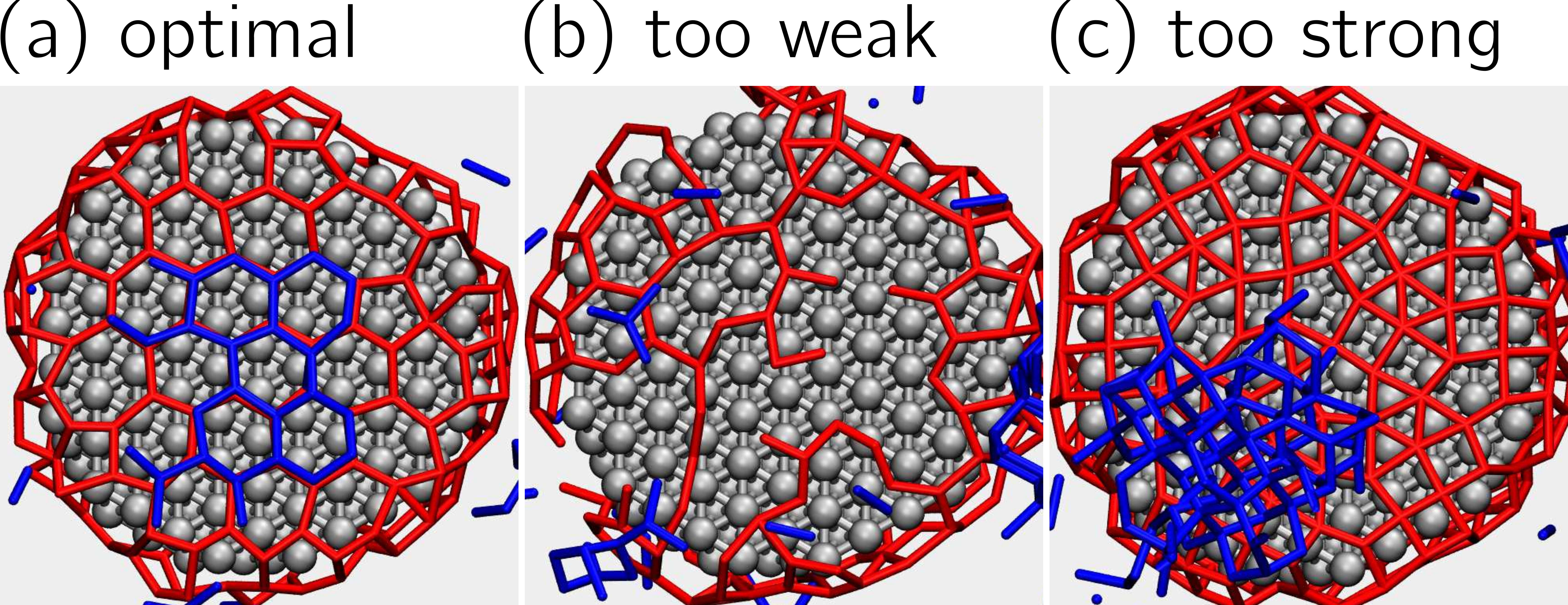}
  \caption{(color online) Sensitivity of the water structure at the
    surface of the NP to the water-NP interaction strength. (a) Ice
    nucleation at the NP with $E_{\trm{ads}}/\Delta H_{\trm{vap}}
    \approx 0.3$. The water molecules in contact with the NP (colored
    red) form an hexagonal layer commensurate with the surface that
    resembles the basal face of ice. The water molecules directly
    above this contact layer (colored blue) also form a similar
    hexagonal structure. (b) The structure of water at the NP with
    $E_{\trm{ads}}/\Delta H_{\trm{vap}} \approx 0.08$. The
    water-surface interaction is too weak to stabilize an ice-like
    structure. (c) The structure of water at the NP with
    $E_{\trm{ads}}/\Delta H_{\trm{vap}} \approx 1.2$. The water
    surface interaction is too strong and the water molecules cannot
    rearrange into an ice-like configuration. In both (b) and (c),
    nucleation occurs away from the surface in a homogeneous manner.}
  \label{fig:snaps2}
\end{figure*}

Here, in the first of a series of two articles, we present results
from molecular dynamics (MD) simulations where we directly probe
heterogeneous ice nucleation in the presence of a face centered cubic
(FCC) nanoparticle (NP). The NP exposes its hexagonal (111) surface as
its principal facet and can therefore act as a template for the
hexagonal basal face of ice. With this NP completely immersed in
water, shown in Fig.~\ref{fig:plots}(a), we perform a series of
studies in which we systematically explore the dependence of the
nucleation rate on the hydrophilicity of the NP. By comparing these
results to reference simulations of homogeneous nucleation we find a
very interesting dependence of the nucleation rate on NP
hydrophilicity; the NP can both promote and inhibit ice nucleation and
exhibits a maximum nucleation rate at intermediate interaction
strengths with the water. By examining the molecular level details of
the nucleation processes at different hydrophilicities of the NP, we
find that the structure in the immediate vicinity of the interface
couples strongly with the nucleation rate. We then use this
understanding of the underlying chemical physics to design an improved
INA. This first article emphasizes how we can use our microscopic
understanding to control ice nucleation. In the second article, we
discuss certain aspects of the mechanism in greater detail, as well as
contextualizing this work with respect to previous studies on surface
hydrophilicity and ice nucleation.

We have used the single site mW potential to model the interactions
between water molecules \cite{molinero:mW-orig}, which allows us to
investigate length- and time-scales inaccessible to ice nucleation
simulations that employ more traditional empirical potentials
\cite{FD:kaolinite}. The NP was modeled as an FCC crystal with a
lattice constant of 0.392\,nm, consisting of 380 atoms. Previous work
has suggested that such a lattice may aid in structuring water into
ice-like arrangements \cite{pccp-gcmc}. The FCC NP was hemispherical
and exposed its (111) face as its primary facet (approximately 2.5\,nm
in diameter). For the interaction between the NP atoms and the water
molecules, the Lennard-Jones potential $U(r) = 4\epsilon \left[
  (\sigma/r)^{12} - (\sigma/r)^{6}\right]$ was used, where $r$ is the
distance between an NP atom and a water molecule. The hydrophilicity
of the NP was controlled by varying $\epsilon$ (a constant value of
$\sigma = 0.234$\,nm was used throughout). As mentioned earlier, Lupi
\etal{} \cite{molinero:het-jacs,molinero:het-jpca} controlled the
hydrophilicity of graphitic surfaces not only in this manner, but also
by introducing hydrophilic species at the surface, and found opposite
trends: we discuss the possible causes of this apparent discrepancy in
more detail in our second paper \cite{layers-paper}. Interactions were
truncated after 0.753\,nm. This setup yielded contact layers at a
height between $0.2$--$0.25$\,nm above the (111) surface of the NP,
which is in reasonable agreement with values obtained from density
functional theory calculations of water at metal surfaces
\cite{angelos:prl-2003,javi:monomer}. We must emphasize, however, that
we are using simplified model surfaces in order to understand possible
general trends that may underlie heterogeneous ice nucleation and that
one must exercise caution in trying to make one-to-one correspondences
with actual surfaces.

All simulations were performed using the \texttt{LAMMPS} simulation
package \cite{lammps} with 2944 mW molecules in a periodic
supercell. Previous simulation studies have suggested that the
critical ice nucleus varies from ${\sim}10$ water molecules at 180\,K
\cite{molinero:cubic} to ${\sim}85$--$265$ at 220\,K
\cite{doye:umbrella,donadio:pccp-ffs} giving us confidence that our
simulations should not be subject to serious finite size
effects. Furthermore, simulations using a slab geometry (approximate
dimensions of 66\,\AA$^{2}$) with 4000 mW molecules confirm that the
conclusions drawn from this work are not affected by changes to the
box size and shape. For each value of the water-NP interaction energy,
16 MD simulations were performed at 205\,K and 1\,bar. Under these
conditions, bulk liquid mW water is still metastable (as opposed to
unstable) \cite{molinero:nature-struct} but undergoes homogeneous
nucleation on a timescale accessible to computer simulation such that
statistically meaningful rates can be obtained. To detect `ice-like'
molecules, we have used the \texttt{CHILL} algorithm of Moore \emph{et
  al.}  \cite{molinero:chill} (As the \texttt{CHILL} algorithm was
designed for bulk homogeneous nucleation, it does not necessarily
capture the full behavior in regions of broken symmetry
i.e. interfaces. Nevertheless, it is useful as a qualitative visual
aid.) By monitoring the potential energy, we are able to determine the
induction time to nucleation for each simulation and thus the
probability $P_{\trm{liq}}(t)$ that a given system remains liquid
after a time $t$ from the start of the simulation. We are able to
determine the ice nucleation rate $R$ by fitting $P_{\trm{liq}}(t) =
\exp\left[-(Rt)^{\gamma}\right]$, where $\gamma > 0$ is also a fitting
parameter. Further details of the fitting procedure and simulation
setup are provided in the Supplemental Material \cite{Note1}. In order
to gauge the effectiveness of the NP as an INA, we have also studied
bulk homogeneous nucleation, using identical settings.

Explicit simulations of heterogeneous ice nucleation have only
recently started to emerge in the literature (see
e.g. Refs.~\onlinecite{FD:kaolinite, molinero:het-jacs,
  molinero:het-jpca, Patey:AgI, doye:AgI, doye:CGhet}) and to enable a
systematic study, we draw conclusions from over 200 successful
nucleation trajectories in total. Fig.~\ref{fig:plots}(b) shows the
dependence of the nucleation rate on the water surface interaction,
the main finding of this study. Specifically, we have plotted
$\log_{10}(R/R_{\trm{hom}})$ vs $E_{\trm{ads}}/\Delta H_{\trm{vap}}$,
where $R_{\trm{hom}}$ is the bulk homogeneous rate and $\Delta
H_{\trm{vap}}$ is the enthalpy of vaporization of bulk mW water
(10.65\,kcal/mol at 298\,K) \cite{molinero:mW-orig}. A rich variety in
the ice nucleating behavior is seen: the NP is seen to both promote
and, surprisingly, inhibit ice formation. We expect this inhibition
effect to be concentration dependent; as the NP concentration becomes
more dilute, we expect the rates to tend to that of homogeneous
nucleation. At low values of $E_{\trm{ads}}$, the heterogeneous
nucleation rate is approximately two times lower than
$R_{\trm{hom}}$. Thus when the particle is very hydrophobic, it tends
to inhibit nucleation. As the water-surface interaction strength
increases so too does the nucleation rate until it reaches a maximum
at $E_{\trm{ads}}/\Delta H_{\trm{vap}} \approx 0.4$ that is nearly 25
times faster than bulk homogeneous nucleation. Beyond the maximum, the
rate steadily decreases until $E_{\trm{ads}}/\Delta H_{\trm{vap}}
\approx 1.0$. Further beyond this, the rate remains roughly constant
and slightly below $R_{\trm{hom}}$.

We now try to understand this intriguing dependence of the nucleation
rate on the hydrophilicity of the NP. To this end, we have examined in
detail the mechanisms by which nucleation occurred on the NP for the
various interaction strengths. As the (111) surface of an FCC crystal
exhibits hexagonal symmetry, one possible mechanism for heterogeneous
ice nucleation is a template effect whereby the molecules in the
contact layer form an hexagonal structure commensurate with the
surface. Fig.~\ref{fig:snaps2}(a) confirms this, where we show a
typical ice nucleation event in the presence of the NP with
$E_{\trm{ads}}/\Delta H_{\trm{vap}} \approx 0.3$ (close to the maximum
rate). Here we can clearly see that the water molecules in contact
with the (111) surface of the NP do indeed form an hexagonal structure
commensurate with the surface that resembles the basal face of ice. We
can also see that the water molecules directly above this contact
layer also form a similar hexagonal structure. The surface is
therefore acting to promote ice nucleation by providing an arrangement
of adsorption sites that resemble the structure of ice, thereby
stabilizing structural fluctuations towards ice-like arrangements in
the liquid.

Now that we have established that the NP acts to promote ice
nucleation by acting as a template for ice, the dependence of the rate
on $E_{\trm{ads}}$ can easily be understood as a competition between
water-water and water-surface interactions. In
Fig.~\ref{fig:snaps2}(b) we show the structure of water in contact
with the NP for $E_{\trm{ads}}/\Delta H_{\trm{vap}} \approx 0.08$ (the
weakest $E_{\trm{ads}}$ investigated, which inhibits ice
nucleation). Clearly, such a weak water-surface interaction is unable
to stabilize ice-like configurations and in fact, ice nucleation is
seen to occur away from the surface in a homogeneous manner. Willard
and Chandler have found that the structure of the interface between
water and a hydrophobic substrate is akin to the liquid-vapor
interface \cite{willard-chandler:hydrophobic-vapor}; a recent
simulation study from Haji-Akbari \etal \cite{debenedetti:ffs} has
found that ice nucleation is disfavored at the liquid-vapor
interface. This appears to be consistent with our observations and
with those of Lupi \etal{} in
Ref.~\onlinecite{molinero:het-jacs}. Fig.~\ref{fig:snaps2}(c), on the
other hand, shows the structure of water in contact with the NP for
$E_{\trm{ads}}/\Delta H_{\trm{vap}} \approx 1.2$. While this NP also
inhibits ice nucleation, it does so for the opposite reason: the
water-surface interaction is too strong, meaning that the water
molecules cannot rearrange to form an ice-like layer at the
surface. It is also clear that the coverage is higher than when ice
forms at the (111) surface, as shown in Fig.~\ref{fig:snaps2}(a) (we
also show this quantitatively in the second paper in this series
\cite{layers-paper}). For this strongly interacting scenario, we also
see that ice forms away from the surface in a homogeneous manner. This
is also consistent with the observations of Reinhardt and Doye
\cite{doye:CGhet} on ice-like surfaces. The observed coupling between
the molecular mechanism and the ice nucleation rate as we change the
surface hydrophilicity is actually rather simple; we now demonstrate
how we can exploit such simplicity to design a surface with improved
ice nucleating efficiency.

\begin{figure}[t]
  \centering
  \includegraphics[width=0.9\linewidth]{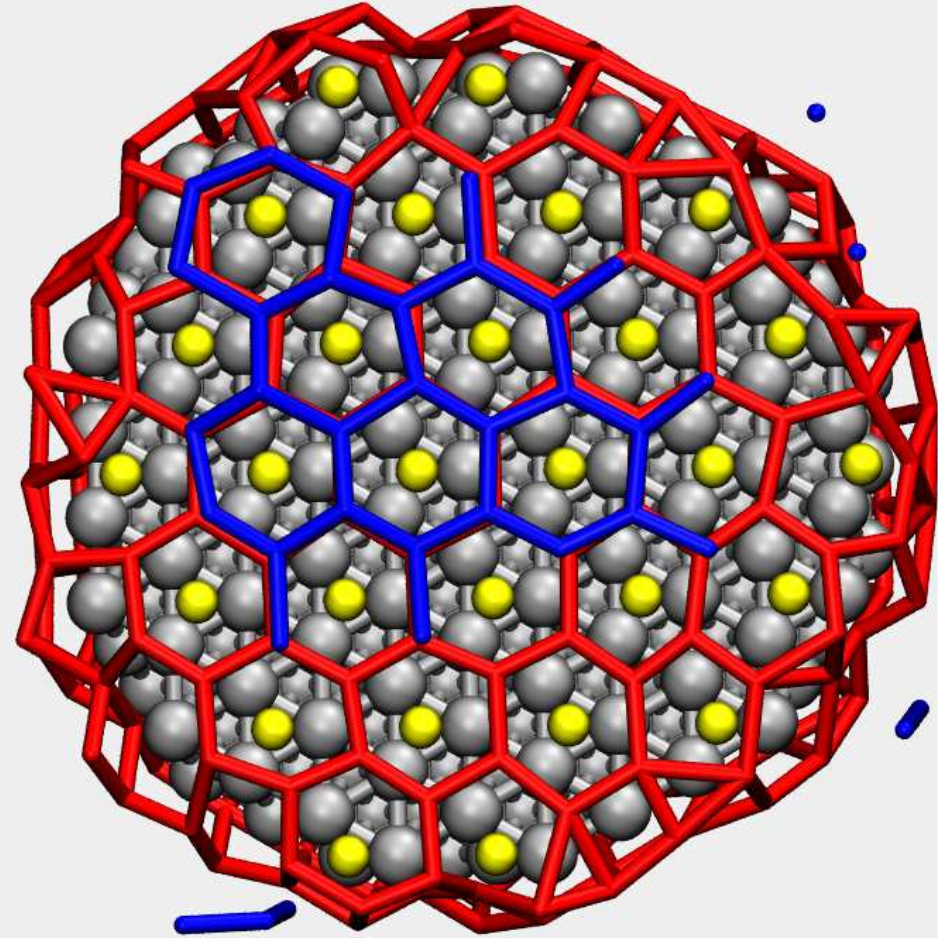}
  \caption{(color online) Surface modification to promote ice
    nucleation for strong water-NP interaction strengths
    ($E_{\trm{ads}}/\Delta H_{\trm{vap}} > 1.0$). By introducing small
    adsorbates (colored yellow) at the (111) surface, the template
    effect can be recovered
    (c.f. Fig.~\protect\ref{fig:snaps2}(c)). The nucleation rate is
    increased by approximately a factor 50 compared to the bare NP for
    $E_{\trm{ads}}/\Delta H_{\trm{vap}} \approx 1.2$, as indicated in
    Fig.~\protect\ref{fig:plots}(b).}
  \label{fig:modified-struct}
\end{figure}

When an ice-like hexagonal overlayer forms at the (111) surface of the
NP, such as in Fig.~\ref{fig:snaps2}(a), it does so with sub-monolayer
coverage i.e. not all of the available adsorption sites are occupied
by water molecules. We refer to these unoccupied adsorption sites on
the (111) terrace as ``excess'' sites. For $E_{\trm{ads}} / \Delta
H_{\trm{vap}} \gtrsim 0.6$, these excess sites are occupied for long
times and for nucleation to occur, an area of decreased coverage at
the surface must occur such that an hexagonal motif can form. This
motif can then act as a template for the hexagonal basal face of ice
(a movie showing this for $E_{\trm{ads}} / \Delta H_{\trm{vap}}
\approx 0.9$ is provided \cite{Note1}). When $E_{\trm{ads}} > \Delta
H_{\trm{vap}}$ it becomes favorable for a water molecule to occupy a
site on the surface, including the excess sites, rather than a
position in the bulk liquid. This prevents the water molecules in the
contact layer from forming the hexagonal arrangements required for ice
nucleation at this NP. By this rationale, if the density of available
adsorption sites was lower, then the template effect (and the enhanced
nucleation rate) may be preserved at higher values of
$E_{\trm{ads}}$. To this end, we have modified the (111) surface of
the NP by adsorbing small molecules, at the excess sites, which only
have a weak interaction with water \cite{Note2}, and recomputed the
nucleation rate with $E_{\trm{ads}} / \Delta H_{\trm{vap}} \approx
1.2$. As seen in Fig.~\ref{fig:modified-struct}, the template effect
is indeed recovered. It is also seen in Fig.~\ref{fig:plots}(b) that
this modified surface enhances ice nucleation by a factor of 50
compared to the unmodified surface. This is a clear demonstration of
how the molecular insight into heterogeneous ice nucleation can be
used to rationally design surfaces of different ice nucleating
ability. Experimentally, this could be realized through adsorption of
small molecules to the surface (e.g. carbon monoxide) or through
surface alloying. In fact, alloying a platinum (111) surface with tin
is observed to promote the formation of an hexagonal ice-like bilayer
under ultra-high vacuum conditions \cite{hodgson:bilayer,
  feibelman:2010:prl}, in a fashion analogous to our modified surface
(the tin atoms at the surface act in part to reduce the density of
adsorption sites). We also note that in our simulations, the
nucleation rate can also be decreased by adsorbing small molecules
such that the NP can no longer act as a template.

More generally, the sensitivity of the nucleation rate on surface
hydrophilicity could be tested by e.g. using nanoparticles of gold or
silica functionalized with organic molecules of varying
hydrophobicity. In addition to using well established methods such as
the droplet freezing techniques \cite{murray:feldspar}, it may also be
possible to exploit recent advances in femtosecond X-ray scattering
techniques that have allowed real-time monitoring of homogeneous ice
nucleation in micron sized water droplets
\cite{nilsson:fancy-xray}. Not only could such an experimental
protocol be used to compare rates of ice nucleation in the presence of
immersed NPs, but information regarding the impact of such NPs on the
microscopic structure of the liquid should also be available.

In summary, we have used computer simulations to systematically
compare heterogeneous ice nucleation rates in the presence of a simple
model nanoparticle of varying hydrophilicity. This complements a
number of recent simulation studies on specific systems
\cite{molinero:het-jacs,
  molinero:het-jpca,FD:kaolinite,Patey:AgI,doye:AgI}. We have seen
that the nanoparticle can promote ice nucleation by acting as a
template for the hexagonal ice lattice, but that the ice nucleating
efficiency is lost if adsorption is too strong, due to a high coverage
of water molecules destroying the template effect. Modification of the
surface such that the coverage of water molecules is reduced recovers
this template effect and enhanced nucleation can be achieved for
strongly adsorbing surfaces, clearly demonstrating how molecular level
understanding of heterogeneous ice nucleation can be used to
manipulate the rate of ice formation. The use of molecular descriptors
to predict useful macroscopic properties of materials has been
successfully used in other fields, such as chemical catalysis
\cite{norskov:NatChem-review}. Designing new catalysts for reactions
such as methanation ($\ce{CO} + \ce{3H2} \longrightarrow \ce{CH4} +
\ce{H2O}$) has relied upon the establishment of a Sabatier principle
based on a computationally tractable quantity (in this case the
dissociation energy of \ce{CO} at the surface)
\cite{norskov:methanation}. We have seen that for the surface
investigated in this study, the adsorption energy of a single water
molecule can be used to describe the heterogeneous ice nucleation
rate. Although a comprehensive set of rules still requires further
experimental and theoretical investigation, the results presented here
suggest that if the surface acts as a template for ice, then one must
tune either the density of adsorption sites, or the propensity of
water to adsorb to the surface. In our second article
\cite{layers-paper}, we show that the variation of the ice nucleation
rate upon surface hydrophilicity is dependent upon the surface
topography, demonstrating that the combined effect of different
surface properties needs to be considered when trying to understand
what makes a good INA. Other properties such as the crystallographic
match to ice and the role of surface defects are also likely to be
important, as will more complex interactions such as electrostatics
and explicit hydrogen bonding. The results presented in this letter
serve as a platform upon which future studies can be conducted.

Dr. Gabriele C. Sosso is thanked for sharing his data from larger
simulations in a slab geometry. We are grateful to the London Centre
for Nanotechnology and UCL Research Computing for computation
resources, and the UK's national high performance computing service
(from which access was obtained via the UK's Material Chemistry
Consortium, EP/F067496). S.M.K. was supported fully by the
U.S. Department of Energy, Office of Basic Energy Sciences, Division
of Chemical Sciences, Geosciences \& Biosciences. Pacific Northwest
National Laboratory (PNNL) is a multiprogram national laboratory
operated for DOE by Battelle. S.J.C. was supported by a student
fellowship funded jointly by UCL and BES. A.M. is supported European
Research Council under the European Union's Seventh Framework
Programme (FP/2007-2013) / ERC Grant Agreement number 616121
(HeteroIce project) and the Royal Society through a Royal Society
Wolfson Research Merit Award.

%

\end{document}